\documentstyle[twocolumn,epsf]{jpsj}

\title{
Magnetic properties of the Hubbard model on three-dimensional lattices: 
fluctuation-exchange and two-particle self-consistent studies
}

\author{
Ryotaro Arita, Shigeki Onoda$^{1}$,
Kazuhiko Kuroki and Hideo Aoki
}
\inst{Department of Physics, University of Tokyo, Hongo, Tokyo
113-0033\\
$^1$Institute for Solid State Physics, University of Tokyo,
Roppoingi, Tokyo 106-8666}

\recdate{\today}

\abst{
The relation between three-dimensional lattice structure and magnetism
in correlated electron systems is explored 
for face centered cubic (FCC), 
body centered cubic (BCC), and simple cubic (SC) lattices.
In particular, we question which lattice structure 
has the strongest tendency toward 
the ferromagnetism or antiferromagnetism. 
We employ the Hubbard model 
to calculate the spin susceptibility and the single-particle spectrum 
with the fluctuation-exchange (FLEX) and the two-particle self-consistent 
(TPSC) approximations in the weak coupling regime. 
We have shown that 
(i) ferromagnetic spin fluctuations become dominant
when the Fermi level lies around a sharp peak (divergence) 
in the density of states ($D(E)$) near the bottom of the band, which occurs 
for FCC with/without next nearest neighbor hoppings ($t'$) 
or BCC with an appropriate value of $t'$.   
Among the cases studied, the ferromagnetic fluctuation is found to be 
the strongest for FCC with a finite $t'$.
(ii) When the peak in $D(E)$ resides around 
the band center as in bipartite SC or BCC, 
antiferromagnetic fluctuations become dominant 
when the band is close to the half-filling, 
with the fluctuation being much stronger in BCC.
}

\kword{three-dimensional Hubbard model, ferromagnetism, antiferromagnetism,
fluctuation exchange approximation, two-particle self-consistent
approximation
}

\begin{document}
\sloppy
\maketitle

\def\runtitle{Magnetic properties of the Hubbard model on three-dimensional
lattice}
\def\runauthor{Ryotaro Arita {\it et al}}

\section{introduction}
The metallic ferromagnetism in repulsively interacting 
electron systems is among central problems in 
solid state physics, but is still some way from a 
complete understanding despite a long history of investigations.  
After the seminal papers by Kanamori~\cite{Kanamori}, 
by Hubbard~\cite{Hubbard63}, and by Gutzwiller~\cite
{Gutzwiller} appeared back in the 1960's, 
the single-band Hubbard model has been studied intensively 
to understand metallic ferromagnetism. 
Although it is a simplest one conceivable, 
the model can still harbor insights
into the fundamental mechanism for an itinerant ferromagnetism 
in correlated electron systems.

Historically, a guiding principle 
for realizing ferromagnetic states 
was provided by Stoner's mean field theory,
which suggests that the ferromagnetic state should be favored 
for large enough interaction and/or density of states (DOS)
at the Fermi level.
Kanamori~\cite{Kanamori} elaborated this argument with 
the T-matrix approximation
to show that the ferromagnetic states are indeed favored when 
a peak in the DOS around the Fermi level 
is embedded around the bottom of the wide band.  
However, for general band filling,
the extent to which the T-matrix approximation is 
valid is not clear. 

As a different avenue along this line, 
rigorous results were recently obtained 
for ferromagnetism on specific lattices having flat bands.
Lieb~\cite{Lieb} has shown that, 
when we have a bipartite lattice with different numbers of 
sublattice sites, the ground state at half filling 
should be ferrimagnetic for arbitrary magnitudes of 
the repulsion, $U$. 
Different numbers of sublattice sites imply that there is a 
flat band(s) in the one-electron band structure, or 
a delta-function density of states.  
Mielke and Tasaki~\cite{MielkeTasaki} then proved the existence of 
ferromagnetic ground states in systems with flat band 
for more general classes of lattices.

Going back to ordinary lattices, one can then 
look for divergences in the DOS 
to realize a ferromagnetic state in the Hubbard model.  
For three-dimensional lattices, the divergence occurs for 
the face centered cubic (FCC) lattice
(where the DOS diverges at the band bottom, 
or above the bottom if we include next-nearest neighbor 
hopping, $t'$) or for the body centered cubic 
(BCC) lattice (where the DOS diverges 
at the center due to the electron-hole symmetry, 
or away from the center if we include an appropriate value of $t'$).  
Our question here is `among the lattice structures 
having sharply peaked (divergent) DOS's which one has 
the strongest tendency toward ferromagnetism?'.  
While the fully polarized ground state is expected to be
favored when the DOS diverges near the bottom of the band,
the tendency toward the ferromagnetic instabilities at
finite temperatures  is a non-trivial problem,
because $T_C$ of a system whose ground state is fully
polarized is not necessarily higher than that of 
a system whose ground state is partially polarized.

As the background for the problem we can first recapitulate how 
the link between the lattice structure and 
the ferromagnetism has been understood 
in terms of the Hubbard model~\cite{Vollhardt}.  
The link may be tested accurately 
in one-dimensional(1D) systems, since 
the density matrix renormalization group (DMRG) method developed 
by White~\cite{White} is applicable in 1D.
Recently, a saturated ferromagnetic ground state has been found
in the Hubbard model on various lattices.  These include 
a trestle chain~\cite{Daul,Daul2,Shimoi} 
which may be regarded as a generalized flat band model in 1D~\cite{PencShiba},
ladder systems~\cite{Liang,Kohno} and 
nearly flat-band models~\cite{Sakamoto}.

In the opposite limit of infinite dimensions, 
exact calculation by means of dynamical mean
field theory (DMFT) becomes feasible, since 
the self energy becomes site-diagonal or {\bf k}-independent.
Using finite-temperature quantum Monte Carlo (QMC) 
techniques, Ulmke~\cite{Ulmke}
has solved the DMFT equations to discuss the stability of metallic 
ferromagnetism for FCC-type lattices in 
infinite dimensions.  
Wegner {\it et al.}~\cite{Nolting97} obtained similar results 
with an iterated perturbation theory instead of QMC. 
If one compares these results with that for the hypercubic lattice by 
Obermeier {\it et al.}~\cite{Obermeier}, FCC is seen to be 
favorable for ferromagnetism.  

For two-dimensional (2D) or three-dimensional (3D) systems, 
which are of realistic interests, the situation is less conclusive,
since we cannot apply the $d=\infty$ result straightforwardly.  
For 2D Hlubina {\it et al.}~\cite{Hlubina,Hlubina2}
suggested a possibility of the ferromagnetic ground state 
when the Fermi level is located near the van Hove singularity 
in the $t$-$t'$ Hubbard model for low densities.
For 3D there are many studies with a variety of approximate methods.
Hanisch {\it et al.}~\cite{Hanisch} studied the instability of Nagaoka's 
saturated ferromagnetic ground state~\cite{Nagaoka} 
for various lattices with a variational method.
The DMFT, which has originally been developed for $d=\infty$, 
has also been applied to 3D by Ulmke~\cite{Ulmke} by plugging 
the DOS of the FCC lattice into 
the DMFT equations to obtain the phase diagram
for FCC. However, the validity 
of the approximation, i.e., 
whether we can neglect the {\bf k}-dependence of the self energy in 3D,
is quite an open question. 
Nolting~\cite{Nolting91} {\it et al.} also studied the stability 
of the ferromagnetic state for the FCC Hubbard model with the spectral
density approach, which may give reasonable results
in the strong coupling limit~\cite{kubo-tada1},
but the obtained $T_C$ turns out to be several times larger than 
those estimated in the DMFT.

Since the above approaches involve one approximation or another 
(such as the strong-coupling limit, infinite dimension, etc), 
further investigations with different approaches should shed a new light
on the problem.  
The purpose of the present paper is to investigate the magnetic properties of
the 3D Hubbard model from the weak-coupling side, where 
the fluctuation-exchange (FLEX) 
and the two-particle self-consistent (TPSC) approximations are 
used to explore ferromagnetism for the first time.  
With these methods we 
look into FCC, BCC, and simple cubic (SC) lattices with or without 
next-nearest neighbor hoppings.  

The FLEX, introduced by Bickers {\it et al.}~\cite{Bickers,Bickers2}, 
takes a set of skeleton diagrams 
for Luttinger-Ward functional to generate
a ($k$-dependent) self energy that is computed self-consistently
based on the idea of Baym and Kadanoff~\cite{Baym,Baym2}.
Hence the FLEX is a kind of self-consistent 
perturbation approximation with respect to $U$.
The method has been applied to the analysis of the 2D Hubbard model
~\cite{Bickers,Bickers2,Dahm,Deisz}
in the context of high-T$_c$ cuprates, 
superconducting ladder systems~\cite{Kontaniladd}
or organic superconductors~\cite{Kontaniorg,Schmalian}.  
The FLEX breaks down in the intermediate to strong
coupling regimes where the vertex corrections become significant,
as seen from the fact that the pseudo-gap or the Hubbard
bands are not reproduced~\cite{Vilk}, but 
the method should give reasonable results at least 
in the weak-coupling regime, namely
as far as the fluctuations do not grow too much.

In Kanamori's theory~\cite{Kanamori}, vertex correction is 
taken into account with the T-matrix approximation, 
which is only valid in the dilute limit.  
Here we have opted for the FLEX approximation, 
since the theory, with no restrictions on the electron density, 
can discuss the density dependence of the spin fluctuation, 
although the vertex corrections are neglected with 
only the self-energy corrections are taken into account there.  
In this context, Moriya and Takahashi~\cite{MoriyaTakahashi} developed a
theory which takes account of both the 
vertex corrections and the self-energy corrections.  
Along this line,
Usami and Moriya~\cite{Usami} discussed the ferromagnetic metals
such as Fe, Co, Ni with a functional integral method.
Similar or more elaborate calculations were done by 
Hasegawa~\cite{Hasegawa} and by Hubbard~\cite{Hubbard81}.
However, these are done at the cost of introducing 
phenomenological parameters, while 
the FLEX can be implemented from a microscopic level 
(e.g., the Hubbard model) 
without introducing any phenomenological parameters.

The TPSC method, on the other hand, takes vertex correction 
into account in a self-consistent manner at the two-body level 
by assuming a parameterized form for the 
Luttinger-Ward functional, where renormalized 
interactions $U_{\rm sp}$ and $U_{\rm ch}$ are introduced.
In term of these `effective $U$' we assume the same functional forms 
(see eqns (\ref{sumrule1},\ref{sumrule2}) below) as
in the RPA for charge/spin susceptibilities.
These parameters are determined so as to satisfy the sum rules or the 
constraints on a two-particle property, {\it e.g.}, the 
double occupancy $\langle n_{\uparrow} n_{\downarrow} \rangle$.
For the 2D Hubbard model on square lattice,
Vilk {\it et al.}~\cite{Vilk} succeeded in reproducing the pseudo-gap and 
the upper/lower Hubbard bands, which indicates that  
the TPSC approach may be valid in the intermediate 
coupling regime as well~\cite{Kyung}.
However, a numerical difficulty can arise in the calculation 
of the chemical potential as we shall see later (\S IIIA), 
so we concentrate on the FCC lattice with
next-nearest neighbor hoppings, for which the ferromagnetic
spin fluctuation turns out to be the strongest among the cases studied 
in FLEX, and is of interest.

Both FLEX and TPSC methods will become invalid in the strong-coupling regime,
while strong repulsion is often thought to be 
required for the realization of 
ferromagnetic ground states as in Nagaoka's case~\cite{Nagaoka}.  
Nevertheless, 
we expect that the tendency toward a ferromagnetic phase
transition in the weak-coupling approach may be captured even when 
the transition itself occurs outside the regime.  
Here we have numerically checked 
to which value of $U$ the method may be relied upon.  
Another purpose of the present paper is to discuss in which 
lattice structure the {\it anti}ferromagnetism is most favored.
Antiferromagnetism, contrary to ferromagnetism, is tractable with 
approaches in the weak-coupling regime in general.

The present paper is then organized as follows.
In section II, after introducing the model,
we review the formalism of FLEX and TPSC.
Section III is devoted to the results and discussions, while 
section IV summarizes the results.

\section{Formulation}
\subsection{Model Hamiltonian}
We consider the conventional single-band Hubbard model,
\begin{equation}
{\cal H}=\sum_{\langle i,j \rangle \sigma} 
t_{ij}c^{\dagger}_{i\sigma}c_{j\sigma}
+U\sum_i n_{i \uparrow}n_{i \downarrow},
\end{equation}
where $c^{\dagger}_{i\sigma}$ creates an electron at the $i$-th site
with spin $\sigma$, $n_{i\sigma}\equiv c^{\dagger}_{i\sigma}c_{i\sigma}$ 
is the number operator.
We take $t_{ij}=t$ for nearest neighbors and $t_{ij}=t'$ for 
second-nearest neighbors.
The energy dispersions for FCC, BCC, and SC lattices are given as
\begin{eqnarray}
\label{FCCDIS}
\varepsilon^{\rm FCC}({\bf k})&=&4t\sum_{i<j}
\cos(k_i)\cos(k_j)+2t'\sum_{i=1}^{3}\cos(2k_i),\\
\label{BCCDIS}
\varepsilon^{\rm BCC}({\bf k})&=&8t
\cos(k_1)\cos(k_2)\cos(k_3)
+2t'\sum_{i=1}^3 \cos(2k_i),\\
\varepsilon^{\rm SC}({\bf k})&=&2t\sum_{i=1}^3 \cos(k_i),
\end{eqnarray}
respectively, where $k_1\equiv k_x$, $k_2\equiv k_y$, $k_3\equiv k_z$.  
To facilitate the calculation here we take a cubic Brillouin zone 
($-\pi <k_i\leq \pi$) by considering two(four) equivalent, 
interpenetrating FCC(BCC) lattices.  
Hereafter, we set $t=1$.

\subsection{The FLEX approximation}
We first review the formalism of the FLEX approximation.
We start from the Luttinger-Ward functional $\Phi$ which is defined as
$\Omega=\Phi+T\sum [\ln G-\Sigma G ]$, where $\Omega$ is the 
thermodynamic potential, $G$ is the dressed Green's function
and $\Sigma$ is the self energy.
The Feynman diagrams for $\Phi$ considered in the FLEX approximation
are shown in Fig. \ref{FLEXLW}.
\begin{figure}
\epsfxsize=8cm
\caption{
The Luttinger-Ward functional considered in the FLEX approximation.
Full lines represent the dressed Green's function 
while broken lines  $U$.
The summation is over 
$n$, the total number of interaction lines.
}
\label{FLEXLW}
\end{figure}

The self energy can be obtained with a functional derivative 
as 
\begin{eqnarray}
\label{Chi}
\Sigma(k)&=&\delta \Phi/\delta G \nonumber \\
&=&\frac{1}{N}
\sum_{q} \left[ G(k-q)V^{(2)}(q) \nonumber \right.\\
&&\left. +G(k-q)V^{({\rm ph})}(q)+G(-k+q)V^{({\rm pp})}(q) \right].
\end{eqnarray}
This consists of the contribution from the bubble diagrams, 
\begin{equation}
V^{(2)}(q) = U^2\chi_{\rm ph}(q),
\end{equation}
the contribution from the particle-hole channel, 
\begin{eqnarray}
V^{({\rm ph})}(q)&=&\frac{1}{2}U^2\chi_{\rm ph}(q) 
\left[ \frac{1}{1+U\chi_{\rm ph}(q)}-1 \right] \nonumber\\ 
\label{PH}
&&+\frac{3}{2}U^2\chi_{\rm ph}(q) 
\left[ \frac{1}{1-U\chi_{\rm ph}(q)}-1 \right], \\
\end{eqnarray}
and the contribution from the particle-particle channel, 
\begin{equation}
V^{({\rm pp})}(q) = -U^2 \chi_{\rm pp}(q)
\left[\frac{1}{1+U\chi_{\rm pp}(q)}-1 \right]
\label{PP}
\end{equation}
where
\begin{eqnarray}
\chi_{\rm ph}(q)&=&-\frac{1}{N}\sum_k G(k+q)G(k), \\
\chi_{\rm pp}(q)&=&\frac{1}{N}\sum_k G(k+q)G(-k).
\end{eqnarray}
Here we have denoted $q\equiv ({\bf q},i\epsilon_\nu)$ and 
$k\equiv ({\bf k},{\rm i}\omega_n)$, 
$\epsilon_\nu=2\pi \nu T$ is the Matsubara frequency for bosons 
while $\omega_n=(2n-1)\pi T$ for fermions, and
$N$ is the total number of sites.

The Dyson equation is written as
\begin{equation}
{G({\bf k},\omega_n)}^{-1} = {G^0({\bf k},\omega_n)}^{-1}
-\Sigma({\bf k},\omega_n),
\label{Dyson}
\end{equation}
where
\begin{equation}
{G^0({\bf k},\omega_n)}
 = \frac{1}{{\rm i}\omega_n+\mu-\varepsilon_{\rm k}^0},
\end{equation}
is the bare Green's function with 
$\varepsilon_{\rm k}^0$ being the energy for the free electrons.

We solve the equations (\ref{Chi}) $\sim$ (\ref{Dyson}) by 
setting the chemical potential $\mu$ so as to fix the density of electrons.
The Green's function is computed by iteration until 
a convergence condition,
\begin{equation}
|G^{(r)}(k)-G^{(r-1)}(k)|/|G^{(r)}(k)|<1.0\times 10^{-4},
\end{equation}
for $G^{(r)}(k)$, the Green's function at the $r$-th iteration,
is attained for all the values of $k$.  
We have checked the convergence by 
taking (i) $16^3$ sites for which 1024 Matsubara
frequencies ($-1023 \pi T \leq \epsilon_{\nu} \leq 1023\pi T)$ 
or 2048 frequencies are considered, and 
(ii) a larger $32^3$ sites with 1024 Matsubara frequencies.

We define the RPA spin susceptibility as 
\[
\chi_{\rm RPA}(q)=\frac{\chi_{\rm ph}(q)}{1-U\chi_{\rm ph}(q)}.
\]
There is no vertex correction in this expression.  
As for the static magnetic susceptibility, 
$\chi(0)$ can be calculated, 
as pointed out by McQueen {\it et al.}~\cite{McQueen},
from the slope Tr$[\sigma_z G]$ versus small homogeneous field $h$ (typically
$0.005t \sim 0.01t$),
where $\sigma_z$ is the $z$-component of the Pauli matrix.
This is equivalent to solving 
an integral equation for the response function
with irreducible vertices comprising particle-particle
and particle-hole bubble chains along with the Aslamazov-Larkin type 
diagrams.

The integrated spectral function, or the DOS, is given by 
\[
\rho(\omega)=-\frac{1}{\pi}\sum_{\bf k} {\rm Im}
G({\bf k},\omega+i\delta).
\]
We obtain $G({\bf k},\omega)$ by a numerical 
analytic continuation from the imaginary-axis data,
$G({\bf k},{\rm i}\omega_n)$, with the Pad\'{e} approximation
~\cite{Vidberg}.

\subsection{The TPSC approximation}
Next we review the TPSC method following the
argument by Vilk and Tremblay~\cite{Vilk}. 
This methods starts from assuming that the Luttinger-Ward 
functional $\Phi$ can
be parameterized by two variables, $\Gamma_{\sigma -\sigma}$ and
$\Gamma_{\sigma\sigma}$, as
\begin{eqnarray}
\Phi&=&\frac{N}{2}\sum_\sigma
\left[
\tilde{G}_\sigma (0^+)\Gamma_{\sigma \sigma} \tilde{G}_\sigma (0^+)
\right. \nonumber\\
&&\left.
+\tilde{G}_\sigma (0^+)\Gamma_{\sigma -\sigma} \tilde{G}_{-\sigma} (0^+)
\right]
\label{TPSCphi}
\end{eqnarray}
with
\begin{eqnarray}
\tilde{G}_\sigma (0^+)=\frac{T}{N}
\sum_{k}\exp(ik0^+)G_\sigma(k).
\end{eqnarray}
If we introduce spin and charge irreducible vertices as
\begin{eqnarray}
U_{\rm sp}&\equiv&\Gamma_{\sigma-\sigma}-\Gamma_{\sigma\sigma}, \\
U_{\rm ch}&\equiv&\Gamma_{\sigma-\sigma}+\Gamma_{\sigma\sigma},
\end{eqnarray}
the charge/spin susceptibility must satisfy the sum rules,
\begin{eqnarray}
\frac{T}{N}\sum_q \chi_{\rm ch}(q)&=&\frac{T}{N}\sum_q
\frac{2\chi_{\rm ph}(q)}
{1+U_{\rm ch}\chi_{\rm ph}(q)} \nonumber \\
\label{sumrule1}
&=&n+2\langle n_{\uparrow}n_{\downarrow} \rangle -n^2 ,\\
\frac{T}{N}\sum_q \chi_{\rm sp}(q)&=&\frac{T}{N}\sum_q
\frac{2\chi_{\rm ph}(q)}
{1-U_{\rm sp}\chi_{\rm ph}(q)} \nonumber \\
&=&n-2\langle n_{\uparrow}n_{\downarrow} \rangle.
\label{sumrule2}
\end{eqnarray}
The RPA-like form of the susceptibilities can be readily derived 
if we assume that 
the irreducible four-point vertices $\Gamma$'s are local in time and space.
Since the self energy corresponding to the trial Luttinger-Ward functional
is constant, the irreducible susceptibility $\chi_{\rm ph}$
coincides with the non-interacting form.
As a first step in the approximation for the chemical potential
$\mu$, we choose the same value as that of the non-interacting 
system, $\mu_0$.

The equations (\ref{sumrule1}) and (\ref{sumrule2}) determine
$U_{\rm sp}$ and $U_{\rm ch}$ as a function of the double occupancy, 
$\langle n_{\uparrow} n_{\downarrow} \rangle$.
If we follow the idea of the local-field approximation for
the electron gas due to Singwi {\it et al.}~\cite{Singwi}, we can assume
a relation,
\begin{eqnarray}
U_{\rm sp}&\equiv& g_{\uparrow \downarrow}(0)U ,\\
g_{\uparrow \downarrow}(0) &\equiv& 
\frac{\langle n_{\uparrow} n_{\downarrow}
\rangle} {\langle n_{\uparrow} \rangle \langle n_{\downarrow} \rangle}.
\end{eqnarray}
from which $U_{\rm sp}$ or $U_{\rm ch}$ can be determined 
self-consistently.

In the next level of the approximation, 
the self energy is assumed to be 
\begin{eqnarray}
\Sigma^{(1)}_\sigma (k) &=& Un_{-\sigma}+ \frac{U}{2}\frac{T}{N} 
\nonumber\\
&& \times \sum_q 
\left[ 
U_{\rm sp}\chi_{\rm sp}(q)+U_{\rm ch}\chi_{\rm ch}(q)\right] 
G^{(0)}_\sigma(k+q),
\end{eqnarray}
for which Green's function is given as
\begin{equation}
G^{(1)}_\sigma(k)
=\frac{1}{{\rm i}\omega_n+\mu_1-\varepsilon_{\rm k}^0-\Sigma^{(1)}_\sigma(k)},
\end{equation}
where $\mu_1$ is obtained by the condition,
\[
n=\frac{T}{N}\sum_k G^{(1)}_\sigma (k)\exp(-{\rm i}\omega_n 0^-).
\]
An important advantage of this expression for
the self energy is that it satisfies the
relation, ${\rm Tr} \Sigma^{(1)}G^{(0)}=2U\langle n_{\uparrow}
n_{\downarrow} \rangle$.  
We can then use the extent to which this relation is numerically 
fulfilled as an internal accuracy check.
Using $G^{(1)}$, Vilk {\it et al.}~\cite{Vilk} succeeded to
reproduce the pseudogaps and the Hubbard bands in the 2D Hubbard
model on square lattice.

When the Luttinger sum rule, 
\[
1/N\sum\theta(-\varepsilon_{\bf k}+\mu_1-\Sigma^{(1)})=n_\sigma ,
\]
is satisfied,
we can take the approximation
$\mu_1-\Sigma({\bf k}_F,0)=\mu_0$.  
Vilk {\it et al.}~\cite{Vilk} suggests that, 
if one considers a region near the half-filling or 
in the strong-coupling regime,
one may improve the approximation by using $\mu_1-\Sigma({\bf k}_F,0)$
in place of $\mu_0$ in calculating the irreducible
susceptibility, $\chi_{\rm ph}$.

\section{Numerical results}
\subsection{FCC lattice}
Now we turn to the present numerical results.  
We first focus on the FCC lattice.
We study the following two cases; (i) a finite $t'=0.5$ with $n=0.2, 0.6, 1.0$
and (ii) nearest-neighbor hopping only ($t'=0$) for the 
same set of band fillings.  
The DOS for the non-interacting case is displayed in Fig. \ref{FCCDOS}(a), 
which is highly asymmetric and has a peak around the bottom of the band.
This is exactly the situation studied by 
Kanamori~\cite{Kanamori} with the T-matrix approximation: 
a dilute case with a large DOS at the Fermi level.

\begin{figure}
\epsfxsize=6cm
\caption{
The density of non-interacting states for
(a) FCC lattice with 
$t'=0$ (solid line) or $t'=0.5$ (dashed line),
(b) BCC lattice with 
$t'=0$ (solid line) or $t'=1.0$ (dashed line), and 
(c) SC lattice.
}
\label{FCCDOS}
\end{figure}

In introducing the interaction, we start with a discussion on 
how strong the Coulomb interaction may be for 
the FLEX approximation to be reliable.
In Fig. \ref{FCCCHI}, we plot the static magnetic susceptibility, 
$\chi\equiv\partial {\rm Tr}[\sigma_z G]/\partial h$, 
and the RPA spin susceptibility, $\chi_{\rm RPA}({\bf 0},0)$, 
as a function of temperature $T$ for $t'=0.5$, $n=0.2$ 
for various values of $U=0.4 \sim 2.0$.
We deliberately took $t'=0.5$ and $n=0.2$ because the 
ferromagnetic spin fluctuation turns out to be 
the strongest for this parameter set
as we shall see below, so that FLEX is expected to be 
least reliable. Indeed, the difference 
between $\chi$ and $\chi_{\rm RPA}$ is seen to be large for $U=2$. 
This means that we cannot approximate the irreducible
effective interaction or the four-point vertex $\Gamma=\delta^2\Phi/
\delta G \delta G$ as $U$ in this case.

\begin{figure}
\epsfxsize=6cm
\caption{
The FLEX static magnetic susceptibility (dashed lines)
and the RPA susceptibility (solid lines)
for the Hubbard model on an FCC lattice with $t'=0.5$ and $n=0.2$
as a function of temperature for $U=2$(a), 
$U=1$(b) and $U=0.4$(c).
}
\label{FCCCHI}
\end{figure}

As a further check, Fig. \ref{FCCDOUBLE} looks at the double occupancy, 
$\langle n_{\uparrow}n_{\downarrow} \rangle$
normalized by $\langle n_{\uparrow} \rangle \langle n_{\downarrow}\rangle$, 
evaluated from the relation,
${\rm Tr} G_\sigma \Sigma_\sigma=
U\langle n_{\uparrow}n_{\downarrow} \rangle$.
We can see that for $U=2$ and $n=0.2$, 
$\langle n_{\uparrow}n_{\downarrow} \rangle$ becomes negative,
which is unphysical.
Hence, for $t'=0.5$ and $n=0.2$, a most stringent case, the FLEX 
approximation
breaks down at least for $U>2$.  Thus we set $U\leq 2$ hereafter.

\begin{figure}
\epsfxsize=6cm
\caption{
A reliability check for the FLEX method from 
the double occupancy $\langle n_{\uparrow}n_{\downarrow} \rangle$
normalized by $\langle n_{\uparrow} \rangle \langle n_{\downarrow}
\rangle$ as a function of $U$ for $t'=0.5$ FCC lattice 
for various values of the band filling, $n$.
}
\label{FCCDOUBLE}
\end{figure}

Now we are in position to present the result for the 
density of states and susceptibility.  
Figure \ref{FCCT05DOS} shows the DOS for $t'=0.5$, $U=1, 2$ for
the density varied from $n=0.2$
(a), $n=0.6$(b) to $n=1.0$(c) with $T=0.1$.
The Fermi level is seen to be right at the peak of the DOS for $n=0.2$, so 
that the ferromagnetic state should be favored 
according to Stoner's mean field picture.

\begin{figure}
\epsfxsize=6cm
\caption{
The density of states (integrated one-particle spectral function) for
$t'=0.5$ FCC lattice with $n=$0.2 (a),
0.6(b), 1.0(c). The solid (dashed) lines are for $U=1 (2)$.  
The insets are blow-ups of the DOS around the
Fermi level. 
}
\label{FCCT05DOS}
\end{figure}

If we then look into the 
RPA FLEX spin susceptibility $\chi_{\rm RPA}({\bf k},0)$ 
in Fig. \ref{FCCT05CHI}, 
the ferromagnetic spin fluctuation is indeed strong 
(i.e., $\chi({\bf k})$ enhanced at $\Gamma$ point) for $n=0.2$ 
as expected from the DOS.  
(Note that the susceptibility takes the same value at 
$\Gamma, {\bf k}=(0,0,0)$ and $K, (\pi,\pi,\pi)$ 
for the convention of taking the FCC dispersion (\ref{FCCDIS}).)

\begin{figure}
\epsfxsize=6cm
\caption{
The RPA spin susceptibility as a function of the wavenumber 
over the Brillouin zone (inset) 
for $t'=0.5$ FCC lattice with $n=$0.2 (a), 0.6(b) or 1.0(c) 
for $U=0$ (solid lines), $U=1$ (dashed lines) or $U=2$ (dotted lines).  
}
\label{FCCT05CHI}
\end{figure}

For an intermediate density $n=0.6$, 
on the other hand, 
the ferromagnetic spin fluctuation disappears.
This is in contrast with the results by Ulmke~\cite{Ulmke},
who predicted, for the strong-coupling regime, 
that the ferromagnetic state is most
favored for intermediate densities, 
and that there is a finite lower critical 
density $n_c=0.1\sim0.2$ below which
the ferromagnetic state is not realized.
For $n=1.0$ (half filling), we cannot find any instabilities
at $(\pi,0,0)$ either, which has been predicted by Ulmke in the 
strong-coupling regime~\cite{Ulmke}.


In Fig. \ref{FCCPHASE}, we show 
$\chi_{\rm RPA}({\bf 0},0)$ for $U=2$ as a function
of $n$. We can see that the ferromagnetic fluctuation is
most dominant for $n=0.2 \sim 0.3$.
If we calculate, for $U=5$, the temperature at which 
$\chi_{\rm RPA}({\bf 0},0)$ diverges as a function of the density $n$,
we find that the temperature takes its maximum value $\sim 0.1t$
for $n\sim 0.2$.
Since the value of $U=5$ is too large for the FLEX approximation to be
valid, at least for $n\sim 0.2$ as mentioned above, 
we cannot naively identify these as the transition temperatures to 
a ferromagnetic state.  
However, the density dependence of these temperatures
suggests that the ferromagnetic spin fluctuation is most dominant
around $n\sim0.2$.
Therefore, we believe that the tendency of the ferromagnetic spin 
fluctuation being most dominant around $n\sim0.2$ 
in the weak coupling regime should be real.

\begin{figure}
\epsfxsize=6cm
\caption{
$\chi_{\rm RPA}(0)$ plotted as a function of $n$ for $U=2$ and $T=0.1$.
}
\label{FCCPHASE}
\end{figure}

Having identified in FLEX that the ferromagnetic
spin fluctuation is the strongest for FCC with $n\sim 0.2$, 
we move on to check whether the strong ferromagnetic spin fluctuation
remains in the TPSC approximation, which 
includes vertex corrections.  
The qualitative features of the spin susceptibility
$\chi$ obtained with TPSC has turned out to be similar to $\chi_{\rm RPA}$ 
obtained with FLEX.  
Here, we focus on how strong the ferromagnetic
spin fluctuation diverges for the case of $n=0.2$ and $t'=0.5$.
We have checked the internal self accuracy, i.e., 
the difference between ${\rm Tr} \Sigma^{(1)}G^{(1)} $
and ${\rm Tr} \Sigma^{(1)}G^{(0)} $~\cite{Vilk},
to find around $10\%$ accuracy at worst. 
To check the convergence of
eq. (\ref{sumrule1}) or eq. (\ref{sumrule2}),
8192 Matsubara frequencies are required,
so we have calculated up to $N=16^3$.

The TPSC result for $1/\chi$ as a function of $T$ is plotted in
Fig. \ref{TPSCCHI} for $U=1 \sim 3$. 
Here the chemical potential $\mu$ is approximated 
to $\mu_0$. The result exhibits a Curie-Weiss like behavior 
($\chi \sim 1/(T+\theta)$) with $\theta > 0$ for $U \leq 3$, 
so that the ferromagnetic transition does not take place as 
far as this result is concerned.  
If we approximate the chemical
potential $\mu$ as $\mu_1-\Sigma({\bf k_F}, 0)$, 
as mentioned at the bottom of \S IIC, 
we can see that the ferromagnetic spin fluctuation
becomes more divergent ($\theta < 0$ for $U = 3$) as seen in the inset. 
For $t'=0.5$ and $n=0.2$, a small change in the chemical potential 
causes large changes in the solution for $U_{\rm sp}$ or $U_{\rm ch}$ 
(eqns(\ref{sumrule1},\ref{sumrule2})).  
However, the tendency detected here that the chemical potential 
decreases as the approximation for 
the chemical potential is improved should be correct, and
it can be shown, at least in the present case of FCC,
that $U_{\rm sp}$ becomes larger for smaller $\mu$.
Therefore, 
although 
$\mu=\mu_1-\Sigma({\bf k}_F, 0)$ suffers from some numerical 
inaccuracies due to an error in 
the Pad\'{e} fit to $\Sigma({\bf k}_F, 0)$, 
the dominance of the ferromagnetic spin fluctuation should be reliable.

\begin{figure}
\epsfxsize=6cm
\caption{
The inverse of spin susceptibility for FCC in the TPSC approximation
as a function of $T$ for $U=1 (\Box)$, $U=2 (+)$ or $U=3 (\times)$, 
where $\mu_0$ is adopted for the chemical potential.  
The inset is a similar plot, where
$\mu_1-\Sigma({\bf k}_F,0)$ is adopted for the chemical potential 
}
\label{TPSCCHI}
\end{figure}

When we turn off the next nearest neighbor hopping ($t'=0$), 
the DOS's obtained by FLEX for $n=0.2$(a),
$n=0.6$(b), and $n=1.0$(c) in Fig. \ref{FCCT0DOS} show that 
the peak around the bottom 
becomes much weaker than that for $t'\neq 0$.
Accordingly, $\chi_{\rm RPA}({\bf k},0)$ 
obtained with FLEX in Fig. \ref{FCCT0CHI} shows that 
the ferromagnetic spin fluctuation at low density ($n\sim 0.2)$ is 
much weaker than that for $t'=0.5$.
We can in fact quantify this from the fact that 
only when $U$ is taken to be as large as 10 (where FLEX
is not reliable) does $1/\chi_{\rm RPA}({\bf 0},0)$ become zero 
for $t'=0$.

\begin{figure}
\epsfxsize=6cm
\caption{
The density of states for
$t'=0$ FCC lattice with $n=$0.2 (a),
0.6(b), 1.0(c). The solid (dashed) lines are for $U=1 (2)$.  
The insets are blow-ups of the DOS around the Fermi level. 
}
\label{FCCT0DOS}
\end{figure}

\begin{figure}
\epsfxsize=6cm
\caption{
The RPA spin susceptibility as a function of the wavenumber 
for $t'=0$ FCC lattice with $n=$0.2 (a), 0.6(b) or 1.0(c) 
for $U=0$ (solid lines), $U=1$ (dashed lines) or $U=2$ (dotted lines).  
}
\label{FCCT0CHI}
\end{figure}

\subsection{BCC lattice}
We next turn to the BCC lattice, a typical bipartite lattice.
We again study the two cases in the presence and absence of 
the next nearest neighbor hopping, $t'$; (i) a finite 
$t'=1.0$ with $n=0.3, 0.7, 1.0$, 
and (ii) nearest neighbor hopping only ($t'=0.0$) with $n=0.2, 0.6, 0.8, 1.0$.
We set $T=0.1$ as before.  
First, the non-interacting DOS, displayed in Fig. \ref{FCCDOS}(b), 
has a peak around the band center, which 
resides exactly at the center for $t'=0.0$, 
since a bipartite lattice has an electron-hole symmetry 
when electrons hop to nearest neighbors only.  
As for the accuracy check, 
$\langle n_{\uparrow}n_{\downarrow}\rangle /
\langle n_{\uparrow}\rangle \langle n_{\downarrow} \rangle$
remains positive for $U\le 2$ for cases (i) and (ii).

We start with the result for a finite $t'=1.0$. 
If we look at the DOS for $n=0.3$(a),
$n=0.7$(b), and $n=1.0$(c) in Fig. \ref{BCCT1DOS}, 
for $n\sim 0.7$ the Fermi level is located around the peak of the
DOS, although a fictitious dip appears due to a 
numerical inaccuracy in the analytic continuation.

\begin{figure}
\epsfxsize=6cm
\caption{
The density of states for
$t'=1.0$ BCC lattice with $n=$0.3 (a),
0.7(b), 1.0(c). The solid (dashed) lines are for $U=1 (2)$.  
The insets are blow-ups of the DOS around the Fermi level. 
}
\label{BCCT1DOS}
\end{figure}

In Fig. \ref{BCCT1CHI}, we plot $\chi_{\rm RPA}$ obtained with FLEX
for $n=0.3$(a), $n=0.7$(b), and $n=1.0$(c). 
Although a finite size effect (slight wiggles) is visible even
when we take $N=32^3$, we believe that 
the overall feature of the spin structure is reliable.
As expected, the ferromagnetic spin fluctuation is dominant
when $E_F$ resides at the peak for $n \sim 0.7$, 
but the enhancement in $\chi$ is much weaker than that for the FCC lattice
with $t'=0.5$ and $n=0.2$.~\cite{commentBCC}
In fact $1/\chi_{\rm RPA}({\bf 0},0)$ 
does not become zero for any value of $U$ for $T=0.1$.
Although the divergence of the peak in the DOS
is as sharp as that of FCC with $t'=0.5$,
the strength of the ferromagnetic spin fluctuation
differs between these cases.
This implies that the peak of the DOS located 
around the bottom is more advantageous for the ferromagnetic spin 
fluctuation than when located around the center.

\begin{figure}
\epsfxsize=6cm
\caption{
The RPA spin susceptibility as a function of the wavenumber 
for $t'=1$ BCC lattice with $n=$0.3 (a), 0.7(b) or 1.0(c) 
for $U=0$ (solid lines), $U=1$ (dashed lines) or $U=2$ (dotted lines).  
}
\label{BCCT1CHI}
\end{figure}

When we turn off $t'$ in the BCC lattice, 
the Fermi level coincides with the peak of the DOS for $n=1.0$ as shown 
in Fig. \ref{BCCT0DOS}.  
Here again,
the analytical continuation with the Pad\'{e} approximation
is unstable for $n\geq 0.6$, causing a dip to appear in the DOS.
In Fig. \ref{BCCT0CHI}, we plot the $\chi_{\rm RPA}$ obtained with FLEX
for $n=0.2 \sim 1.0$.  This time the 
spin fluctuation becomes dominant at 
the {\it anti}ferromagnetic point (X point with 
${\bf k}=(\pi,\pi,\pi))$ for $n=0.8$ and $n=1.0$.
For $n=1.0$, $1/\chi_{\rm RPA}({\bf 0},0)$ becomes zero for $U\sim 2$ 
and $T=0.1$. As we shall see below, the 
antiferromagnetic fluctuation is much stronger than
that for the simple cubic lattice.

\begin{figure}
\epsfxsize=6cm
\caption{
The density of states for
$t'=0$ BCC lattice with $n=$0.2 (a),
0.6(b), 0.8(c) or 1.0(d). The solid (dashed) lines are for $U=1 (2)$.  
}
\label{BCCT0DOS}
\end{figure}

\begin{figure}
\epsfxsize=6cm
\caption{
The RPA spin susceptibility as a function of the wavenumber 
for $t'=0$ BCC lattice with $n=$0.2 (a),
0.6(b), 0.8(c) or 1.0(d) 
for $U=0$ (solid lines), $U=1$ (dashed lines) or $U=2$ (dotted lines).  
}
\label{BCCT0CHI}
\end{figure}

\subsection{Simple cubic lattice}
We finally turn to the SC lattice.
We plot the DOS in Fig. \ref{SIMPLEDOS} for
$n=0.2$, $n=0.6$, $n=1.0$, while 
the DOS for the non-interacting case is shown in Fig. \ref{FCCDOS}(c).
They are electron-hole symmetric since SC lattice is bipartite.  
We set $T=0.1$.
Here again $\langle n_{\uparrow}n_{\downarrow}\rangle /
\langle n_{\uparrow}\rangle \langle n_{\downarrow} \rangle$
remains positive for $U\le 2$.  
If we look at $\chi_{\rm RPA}$ obtained with FLEX in  
Fig. \ref{SIMPLECHI}, the ferromagnetic fluctuation is seen to be suppressed.
The only enhancement appears for $n=1.0$ 
at the antiferromagnetic point (K point), but the fluctuation 
is weaker than that for the BCC lattice above.

\begin{figure}
\epsfxsize=6cm
\caption{
The density of states for
$t'=0$ SC lattice with $n=$0.2 (a),
0.6(b), or 1.0(c). The solid (dashed) lines are for $U=1 (2)$.  
}
\label{SIMPLEDOS}
\end{figure}

\begin{figure}
\epsfxsize=6cm
\caption{
The RPA spin susceptibility as a function of the wavenumber 
for $t'=0$ SC lattice with $n=$0.2 (a), 0.6(b) or 1.0(c) 
for $U=0$ (solid lines), $U=1$ (dashed lines) or $U=2$ (dotted lines).  
}
\label{SIMPLECHI}
\end{figure}

\section{Discussions and Summary}
We have studied the Hubbard model on 3D FCC, BCC, and SC lattices
with the FLEX and the TPSC approximations in a weak-coupling regime.
We have calculated the spin susceptibility as well as 
the density of states (integrated single-particle spectral function).
For cases in which the DOS is divergent near the bottom of the band
and the Fermi level is located around the divergence, 
the {\it ferromagnetic} instability is found to be dominant.
Such a situation is realized for 
a low band filling in the FCC lattice with a finite 
next-nearest neighbor hopping ($t'$) or 
close to the half-filling in the BCC lattice with a finite $t'$.
For bipartite lattices 
(SC and BCC lattices with nearest neighbor hopping only), 
by contrast, the {\it antiferromagnetic} fluctuation
is dominant around the half-filling.

We have found, among the cases studied, that the strongest 
ferromagnetic fluctuation occurs for a 
low band filling FCC lattice
with $t'$, 
whose DOS is characterized by the sharpest divergence that 
resides around the band bottom.  
From a view of ``frustration" we may say the following.  
In the absence of $t'$, FCC is frustrated in that 
its constituent plaquette is a tetrahedron where 
antiferromagnetic correlations have to interfere, 
while BCC is bipartite compatible with antiferromagnetism.  
Thus they have different situations, before $t'$ is turned on 
to make the DOS for BCC divergent.  
While both of the divergent DOS and the frustration 
can favor ferromagnetism intuitively, we cannot at present stage 
distinguish effects of the two.   
We can still observe that the present result 
for a stronger ferromagnetic fluctuation in FCC than in BCC
implies that the peak of the DOS lying 
around the bottom of the band may be a key factor.  
As stressed in Introduction, such a comparison could not 
have been performed from Kanamori's theory, which focuses on dilute cases.
For ground state, Hanisch {\it et al.}~\cite{Hanisch}
suggested that the fully-polarized ground state is
favored when the DOS diverges near the
bottom of the band.
On the other hand, we have discussed here
the tendency toward ferromagnetic instabilities
at the finite temperatures, which includes the possibilities
of not only fully-polarized state but also partially polarized state.
Hence the relation between the results of Hanisch {\it et al.}
and our results is not trivial and is an interesting 
future problem.

The discussion on the position of the divergence in DOS and frustration 
reminds us of a situation in 1D, in a quite different context.  
In the flat-band ferromagnetism mentioned in 
Introduction, we can consider quasi-one-dimensional systems in 
which flat bands are realized.  There, flat bands can arise 
in non-frustrated models (such as a chain of squares)~\cite{AritaYajima}, 
but the flat band resides in the middle of the band, so that 
the ferromagnetism is shown to be destroyed when $U$ becomes too large. 
In frustrated models (such as a chain of triangles), by contrast, 
the flat band is situated at the bottom, and the ferromagnetism is 
guaranteed for $0<U<\infty$ from the Mielke-Tasaki theorem.   
Although this observation concerns quite a distinct class of systems, 
further discussions may prove fruitful. 

The FLEX result for the FCC lattice indeed shows that 
the ferromagnetic instability is most dominant in the low-density region.  
This reminds us that Ni has an FCC structure with a low hole concentration 
on the one hand.  
On the other, the result is contrasted with a DMFT result 
for the strong-coupling regime by Ulmke~\cite{Ulmke}, 
which suggests that for FCC lattice with $t' \neq 0$,
the ferromagnetic region 
only appears above $n_c=0.1\sim0.2$ and is peaked at  
an intermediate density ($n \sim 0.6$).  
It is an interesting future problem to 
explore how these two regimes are interpolated.

The antiferromagnetic fluctuation, by contrast, appears for bipartite (i.e., electron-hole symmetric) lattices (BCC or SC), 
and the fluctuation is the strongest for the half-filled BCC with 
a divergent DOS at the Fermi level (for $t'=0$) 
of all the situations surveyed here. We note that it is stronger
than that in the half-filled SC lattice whose DOS has only a finite peak.
This cannot be understood from $\chi_{\rm ph}(\pi,\pi,\pi)$ 
for non-interacting electrons, since there is not much difference 
in this quantity between the two cases.

\section*{Acknowledgment}
R. A. would like to thank S. Koikegami and T. Mutou for
illuminating discussions.
R. A. and S. O. would like to thank H. Kontani for 
discussions on the FLEX.
We would like to thank K. Kusakabe for useful comments.
K. K. acknowledges a Grant-in-Aid for Scientific
Research from the Ministry of Education of Japan.
R. A. is supported by a JSPS Research Fellowship for
Young Scientists.
Numerical calculations were done on SR2201 at the
Computer Center, University of Tokyo.


\begin{thebibliography}{99}
\bibitem{Kyung}
In cond-mat/9802113,9802129,
B. Kyung has argued that
he has obtained the Hubbard bands along this line 
in the strong-coupling regime ($U=8$) as well.
\bibitem{commentBCC}Note that the susceptibility takes the same value at 
$(0,0,0)$ and $(\pi,\pi,0)$ if we take the dispersion eq.
(\ref{BCCDIS}) for the BCC lattice.
\end{thebibliography}
\end{document}